\documentclass[11pt,twoside]{article}

\usepackage{asp2004}
\usepackage{epsf}
\usepackage{epsfig}
\usepackage{lscape}

\begin{document}
\title{Using {\it Spitzer}  to probe the nature of submillimetre galaxies in GOODS-N}  
\author{Alexandra Pope, Douglas Scott, Mark Dickinson, Ranga-Ram Chary, Glenn Morrison, Colin Borys, Anna Sajina} 
\affil{UBC, UBC, NOAO, SSC, UHawaii-IfA/CFHT, Caltech, SSC }   

\begin{abstract}
How does the submm galaxy population detected by SCUBA fit into galaxy evolution? How do these rare starbursting systems, which contribute significantly to high redshift star-formation, relate to other galaxy populations? Deep radio observations have been most useful for studying these systems, but still leave a significant fraction of the population unidentified. Now with the deep IRAC and MIPS images from the GOODS {\it Spitzer}  Legacy program and a re-analysis of the deep radio data, we are able to identify counterparts for a large fraction of SCUBA galaxies in GOODS-N. All of these counterparts are detected by {\it Spitzer} . Given the vast multi-wavelength data in this field, we can study the spectral energy distributions (SEDs) of these systems and determine what is fueling their intense infrared luminosities. A rest-frame composite optical-to-radio SED for all spectroscopically identified submm sources shows that the average SCUBA galaxy is consistent with models of ultraluminous starburst galaxies, although cooler than those observed locally. Because of this, the submm flux alone consistently overestimates $L_{\rm{IR}}$ when using spectral templates which obey the local ULIRG temperature-luminosity relation. The wide range of 24/850 micron flux ratios as a function of redshift indicates the presence of strong mid-IR features, to be confirmed with deep IRS spectroscopy. The IRAC colours of the submm systems provide useful redshift constraints, since, at these redshifts, IRAC samples the stellar bump. The {\it Spitzer}  photometry of this large sample of submm galaxies has allowed us to put constraints on many of the outstanding issues in submm astronomy.
\end{abstract}


\section{Introduction} 
The Submillimetre Common User Bolometer Array (SCUBA, Holland et al.~1999) on the James Clerk Maxwell Telescope (JCMT) is efficient at finding high redshift dusty galaxies in blank-field extragalactic surveys. These galaxies are Ultraluminous, and are known to contribute significantly to the star-formation budget at high redshift (e.g.~Lilly et al.~1999; Chapman et al.~2005). However a general understanding of the submm population is severely limited by the small number of known SCUBA sources, the large beam size at 850$\,\mu$m,  and their faintness at other wavelengths.  
Enormous progress has been made using radio observations to localize the submm sources (e.g.~Ivison et al.~2000; Smail et al.~2000; Barger et al.~2000). At least $1/3$ of currently known 850$\,\mu$m galaxies have no radio counterpart in some of the deepest radio observations and therefore deep observations at other wavelengths are crucial. No other SCUBA field has the extensive multiwavelength coverage of the Great Observatories Origins Deep Survey (GOODS) Northern field. In addition to the VLA radio imaging (Richards 2000; Morrison et al.~in prep), deep {\it Hubble Space Telescope} ({\sl HST\/}) imaging in four bands (Giavalisco et al.~2004) and the {\it Chandra} 2 Msec image (Alexander et al.~2003), GOODS boasts the deepest observations taken in the infrared with IRAC and MIPS on the {\it Spitzer Space Telescope\/} (Dickinson et al.~in preparation). GOODS-N is also home to a large amount of SCUBA data taken by several groups (Hughes et al.~1998; Barger, Cowie, \& Richards 2000; Borys et al.~2003), which have been combined into one large 850$\,\mu$m `super-map' (Borys et al.~2003, 2004; Pope et al.~2005). 

\section{Identifications} 

\begin{figure}[t]
\begin{center}
\includegraphics[width=3.0in,angle=0]{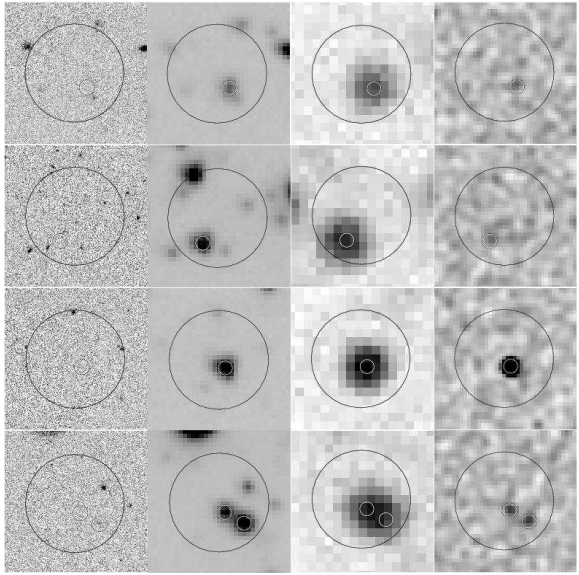}
\caption{Multi-wavelength images of four submm counterparts, from left to right: ACS $z_{\rm{850}}$; IRAC $3.6\,\mu$m; MIPS $24\,\mu$m; and radio $1.4\,$GHz. All images are $20\times20$ arcsec and the black circle is centred on the SCUBA position with a 7 arcsec radius. The smaller black/white circle indicates the submm counterpart. Submm galaxies are quite faint at other wavelengths, but deep MIPS data are extremely valuable for identification.
}
\label{fig:postage}
\end{center}
\end{figure}

We have identified 35 secure $850\,\mu$m sources in the GOODS-N submm super-map (Borys et al.~2003; Pope et al.~2005).
Using all multi-wavelength data, unique counterparts are found for 94 per cent (33 out of 35) of these submm sources. 
26 out of 35 are detected above $3\sigma$ in the new Morrison et al.~reduction of the 1.4$\,$GHz data and {\it all} of the counterparts are detected with IRAC and MIPS. Fig.~\ref{fig:postage} shows multi-wavelength images of four of our submm sources. 

\section{MIPS 24$\,\mu$m properties}   

\begin{figure}[t]
\begin{center}
\includegraphics[width=5.0in,angle=0]{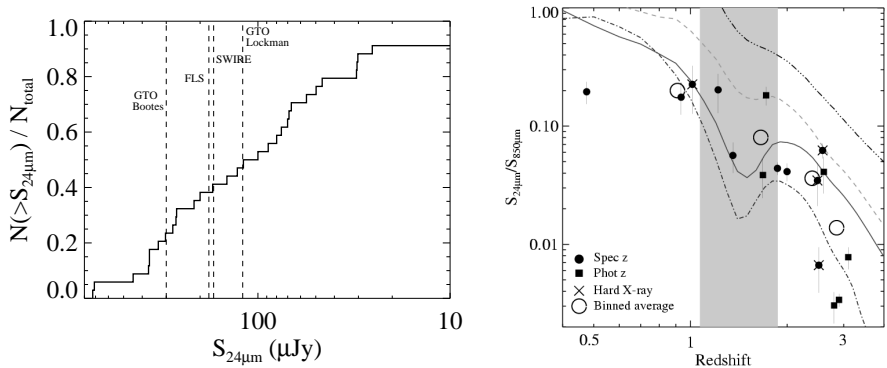}
\caption{
Left: Cumulative distribution of $24\,\mu$m flux for submm counterparts in GOODS-N. 
Right: The $S_{24}/S_{850}$ ratio as a function of redshift for submm sources in GOODS-N.
The dashed curve is an SED template from Chary \& Elbaz 2001 (CE01) models, the solid curve is a modified CE01 template which includes additional extinction from Draine (2003), the dash-dot curve is the observed SED of Arp220 and the dash-dot-dot curve is Mrk231, an IR-luminous AGN. The shaded region represents the region where the 9.7$\,\mu$m silicate feature passes through the 24$\,\mu$m passband.
}
\label{fig:comb1}
\end{center}
\end{figure}

Both SCUBA and {\it Spitzer\/} are able to select distant star-forming galaxies which contribute significantly to the cosmic infrared background.
In blank field surveys, SCUBA is primarily sensitive to the thermal dust emission from ultra-luminous galaxies at high redshift. While {\it Spitzer\/} also detects these galaxies,  its sensitivity is such that it detects galaxies over a wider range of luminosity and galaxy type.
The left panel of Fig.~\ref{fig:comb1} shows the MIPS flux distribution for our $850\,\mu$m selected sample. The vertical dashed lines indicate the 3$\sigma$ limits of several other {\it Spitzer} surveys: GTO shallow survey of the Bootes field (E.~Le Floc'h priv.~communication), FLS, SWIRE (D.~Farrah priv.~communication) and the GTO deep survey of the Lockman Hole East region (Egami et al.~2004).
The GTO shallow survey of the Bootes field is deep enough to detect 25 per cent of the SCUBA galaxies.
With the MIPS depth of the First Look Survey (FLS\footnote{http://ssc.spitzer.caltech.edu/fls/extragal/spitzer.html}), or SWIRE (Lonsdale et al.~2004), we would detect 40 per cent of our submm counterparts. 

In the right panel of Fig.~\ref{fig:comb1}, the $S_{24}/S_{850}$ ratio is plotted as a function of redshift for the sample of submm sources in GOODS-N as compared to several models.
The generally lower $S_{24}/S_{850}$ ratios for the submm sources suggest that they have higher levels of extinction, especially in the mid-IR, than the local ULIRGs which were been used to construct these models. In particular, the silicate absorption feature at 9.7$\,\mu$m may be attenuating the 24$\,\mu$m flux.

\section{Spectral energy distribution}   

\begin{figure}[t]
\begin{center}
\includegraphics[width=5.0in,angle=0]{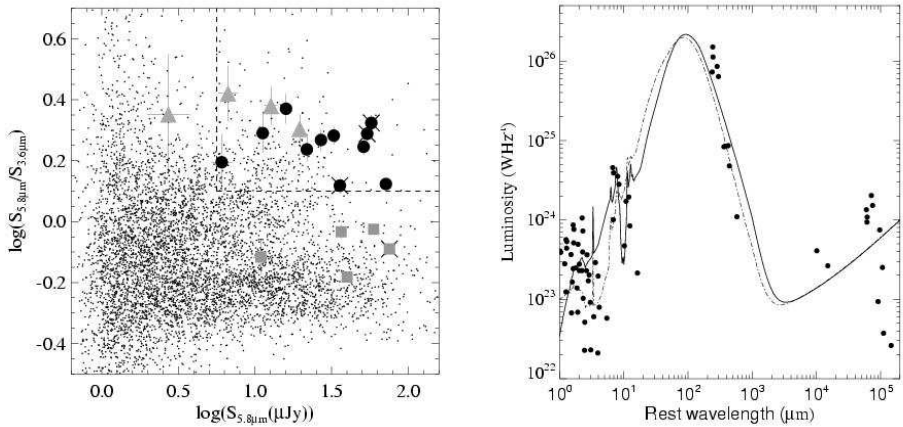}
\caption{Left: IRAC-based colour-magnitude diagram for GOODS-N. Larger symbols are our submm sources, while the small black dots are field galaxies.
The squares are at low redshift ($z<1.5$), the circles are at high redshift ($z>1.5$) and the triangles have unknown redshift. Crosses are sources which contain an AGN, as indicated by the presence of hard X-rays.
Right: Composite SED for the GOODS-N submm sources with spectroscopic redshifts. IRAC through radio photometry data are plotted. 
The solid curve is a modified CE01 model for a cool starburst galaxy, where we have applied additional extinction using the Draine (2003) models to minimize scatter relative to the data points. 
The dash-dot curve is a CE01 model with the same luminosity but a different peak wavelength. This template does not have any additional extinction and is a poor fit to the mid-IR and submm data points.
}
\label{fig:comb2}
\end{center}
\end{figure}

At the redshifts of these submm galaxies, the IRAC channels are sampling the rest-frame near-IR, which is sensitive to direct emission from stars. If these galaxies are not dominated by an AGN at these wavelengths then the IRAC colours should show an indication of the $1.6\,\mu$m local peak of the $f_\nu$ stellar SED.  In the left panel of Fig.~\ref{fig:comb2}, we plot an IRAC-based colour-magnitide diagram, specifically $S_{5.8}/S_{3.6}$ as a function of $S_{5.8}$, for the submm sources and the GOODS-N field galaxies. The redshift of the submm source is indicated by the symbol type. We see that $z\sim1.5$ provides a clear transition in this IRAC colour for the submm sources, due to the filters moving over the stellar peak. The sources which lack redshift information are all consistent with the colours of the higher redshift submm sources. There seems to be no strong colour distinction between the sources with and without a hard X-ray detection, although the former are generally brighter at $5.8\,\mu$m. 

The right panel of Fig.~\ref{fig:comb2} shows a composite SED for the submm sources in our sample which have spectroscopic redshifts, as compared to several models. 
The main conclusion to be drawn from this plot is that high redshift submm galaxies appear to have different SED shapes from those observed locally for galaxies of similar luminosities. A major contribution to this difference could be the temperature of the dust, but it is important to realize that this just means an SED peaking at longer wavelengths. The best-fit submm SED (solid curve in Fig.~\ref{fig:comb2}) peaks at $\sim100\,\mu$m (T$\,\simeq$$\,30\,$K), while a typical local ULIRG template from CE01 (dash-dot curve in right panel of Fig.~\ref{fig:comb2}) of the same luminosity peaks at $\sim85\,\mu$m (T$\,\simeq$$\,34\,$K). 
This result emphasizes the strong selection effects, both locally and at high redshift. Our knowledge of local ULIRGs is not immune to selection effects, since it is dominated by results from IRAS which may preferentially select galaxies with warmer SEDs (peaking at shorter wavelengths). Similarly at high redshift, 850$\,\mu$m imaging preferentially selects galaxies which peak at longer wavelengths, and therefore comparing the two samples becomes problematic.

\section{Infrared luminosities}   

\begin{figure}[t]
\begin{center}
\includegraphics[width=2.8in,angle=0]{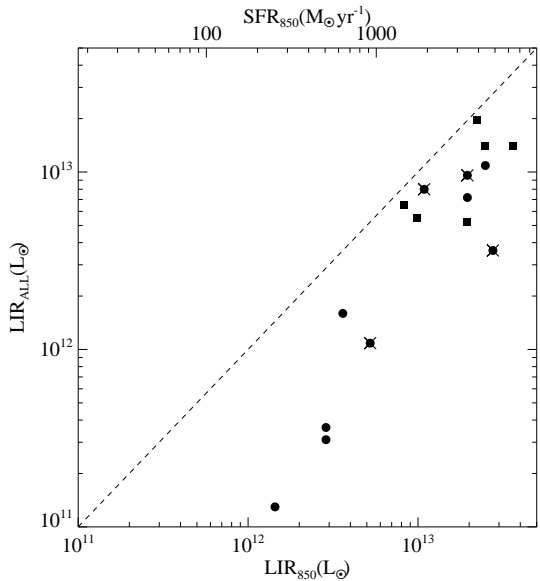}
\caption{Infrared luminosities and SFRs for submm sources.  The bottom axis show the luminosities estimated using only the $850\,\mu$m flux and the redshift. The left-hand axis is the estimated IR luminosity using data at 24$\,\mu$m, $850\,\mu$m and $1.4\,$GHz.
}
\label{fig:lum}
\end{center}
\end{figure}

Using the CE01 models, we have estimated the IR luminosity by fitting to the templates: 1) the $850\,\mu$m flux and the redshift, and 2) the $24\,\mu$m, $850\,\mu$m and $1.4\,$GHz flux and the redshift.
Fig.~\ref{fig:lum} compares the infrared luminosity (LIR) as estimated these two ways, by integrating from 8--1000$\,\mu$m under the best-fit modified CE01 template. 
Filled circles and squares denote submm sources with spectroscopic and photometric redshifts, respectively. The crosses denote sources which contain an AGN, as indicated by the presence of hard X-rays. The top axis shows the corresponding star formation rate (SFR) using the expression in Kennicutt (1998).
Using templates based on local ULIRGs with a fixed luminosity-temperature relation (CE01), the submm flux will typically overestimate the luminosity. This effect is most dramatic for less luminous submm objects which tend to lie at lower redshifts where the 850$\,\mu$m window samples even further from the IR SED peak. 

The median IR luminosity (using radio, submm and infrared flux) for our sample is L$_{\rm{8-1000\,\mu m}}\,$=$\,6.0\times10^{12}\rm{L}_{\sun}$, which corresponds to a SFR of $1100\,$M$_{\sun}\rm{yr}^{-1}$.

\section{Summary}
 Using the deep {\it Spitzer} Legacy images and a re-reduction of the VLA radio data, we find  secure counterparts for a large sample of submm galaxies in GOODS-N. 
The 24$\,\mu$m to 850$\,\mu$m flux density ratio as a function of redshift is low compared to models, suggesting that the silicate absorption feature at 9.7$\,\mu$m may be attenuating the 24$\,\mu$m flux.
A composite rest-frame SED shows that the submm sources peak at longer wavelengths than those of local ULIRGs. This demonstrates the strong selection effects, both locally and at high redshift, which may lead to an incomplete census of the ULIRG population.
We determine the IR luminosity $L_{\rm{IR}}(8-1000\mu$m) by fitting models and find that the submm flux alone consistently overestimates $L_{\rm{IR}}$ when using spectral templates which obey the local ULIRG temperature-luminosity relation.
The shorter {\it Spitzer} wavelengths sample the stellar bump at the redshifts of the submm sources, and we find that the {\it Spitzer} photometry alone provides a model independent estimate of the redshift. 
More details on this work will be presented in Pope et al.~(2006).

\acknowledgements 
AP would like to thank the conference organizers for the opportunity to present this work.
This work was supported by the Natural Sciences and Engineering Research Council of Canada. 
Support for GOODS, part of the {\it Spitzer Space Telescope} Legacy Science Program, was provided by NASA through Contract Number
1224666 issued by JPL, Caltech, under NASA contract 1407.


\end{document}